\documentclass[acmsmall,nonacm]{acmart}

\microtypesetup{expansion=false}

\AtBeginDocument{%
  }

\usepackage{verbatim}

\begin{document}

\title{Why Software Engineering Is Indispensable in the Age of Coding Agents}

\author{Alfonso Fuggetta}
\orcid{0000-0002-9514-0780}
\affiliation{%
  \institution{Politecnico di Milano}
  \department{Dipartimento di Elettronica, Informazione e Bioingegneria}
  \city{Milan}
  \country{Italy}
}
\email{alfonso.fuggetta@polimi.it}

\begin{abstract}
Can AI make Software Engineering (SE) --- the discipline --- obsolete? And can it make software engineers --- the professionals --- redundant? This paper argues that the rise of capable AI coding agents makes SE and software engineers essential, not obsolete: the missing foundation without which AI-assisted development produces misleadingly plausible, unverifiable, and ultimately untrustworthy software. Three structural properties of large language models (probabilistic generation, agnosticism, and semantic statelessness) create a structural vacuum that no amount of training can eliminate. Filling it requires four knowledge levers: methodological knowledge, domain knowledge, design choices, and process choices. All four must be reified as persistent artifacts, and each requires the software engineer as methodologist, mediator, and custodian.
\end{abstract}


\maketitle

\begingroup
\renewcommand\thefootnote{}
\footnotetext{\textbf{Author's version.} Accepted for publication in \textit{Communications of the ACM} (manuscript CACM-26-07-6124). This is the author's final version of the work, submitted to the publisher on 11 October 2026. The definitive Version of Record will be published in \textit{Communications of the ACM}, \url{https://doi.org/10.1145/3858606}; please cite the published version. \copyright{} 2026 Copyright held by the author. Licensed under a Creative Commons Attribution-NonCommercial-NoDerivatives 4.0 International License (CC BY-NC-ND 4.0).}
\addtocounter{footnote}{-1}
\endgroup

\section{What We See}

The advent of coding agents has raised legitimate questions about the fate of Software Engineering and software engineers; in these pages, Welsh went as far as announcing \textit{The End of Programming}. To answer them, one may assess empirical data or practitioners' judgment. The first evidence to enable careful judgment is now emerging: field observation and professional practice, randomized trials, cross-country firm surveys.

Meyer~\cite{meyer2026}, writing in these pages, frames the current moment. He distinguishes three software categories (\textit{acute}, \textit{business}, \textit{casual}) and argues that \textit{vibe coding} --- Karpathy's term for code written by those who ``fully give in to the vibes'' rather than engineering it --- works for the third but not for the first two, where correctness is paramount and coding is only 10--20\% of project cost. He names the \textit{hallucination loop}: a plausible suggestion that sends the developer down a path that will not close. His conclusion: for professional software, AI-assisted development must be paired with formal specification and verification, the move from \textit{probable} to \textit{provable}.

The empirical picture is not uniform. Cui et al.~\cite{cui2025} found in field experiments a 26\% increase in completed tasks across 4,867 developers at Microsoft, Accenture, and a Fortune 100 firm, with the largest gains among less experienced developers; a \textit{Science} study of 160,097 GitHub developers finds the opposite, with gains concentrated among senior developers and no significant effect for early-career ones~\cite{daniotti2026}. Becker et al.~\cite{becker2025} ran a randomized controlled trial with sixteen experienced open-source developers on 246 real tasks in their own mature repositories: tasks on which AI was allowed took 19\% longer, and afterward developers believed AI had helped; a 2026 follow-up by the same team considers speedups now likely, but deems its new measurements unreliable. Xu et al.~\cite{xu2025} show where the cost lands: in a difference-in-differences study of 2,755 open-source repositories, lines of code added rose 17.7\% after Copilot introduction, but core contributors --- those with the deepest ownership of the codebase --- saw a 19\% decline in original code output while reviewing 6.5\% more code and absorbing the added rework. Two difference-in-differences studies presented at MSR 2026, one on an AI IDE and one on coding agents, find the same pattern: quality costs that persist even when velocity gains fade, with static-analysis warnings up 30\% and about 18\%, and code complexity up 41\% and about 39\%~\cite{agarwal2026,he2026}. Time saved on generation is recaptured on supervision, spent on output that looks plausible but requires an external reference for correctness. That reference lives with the developer, not with the tool.

Huang et al.~\cite{huang2025} studied thirteen developers in the field and surveyed ninety-nine more. A distinctive divide has emerged: some developers practice vibe coding, treating interaction flow as evidence of correctness. Professionals do not: they plan, validate output, and retain architectural control, treating the agent as an executor. Where AI-assisted development succeeds, something structured is compensating for what the agent alone cannot do.

At scale, the Stack Overflow 2026 Developer Survey --- 30,903 respondents across 169 countries --- captures the same ambivalence: 83\% use AI tools at work, yet only 6.6\% trust their output for important work decisions, while 48\% trust it when they can easily verify it.

AI coding agents have a structural vacuum: on their own, they lack what they need to succeed. The vacuum follows from what AI tools structurally are; the nature of software makes it consequential, and the state of SE decides whether it is filled. Software is intangible and, as Balzer~\cite{balzer1991} observed, ``rarely consistent'': requirements evolve, knowledge arrives incrementally, practices carry inconsistencies that are features, not errors. Classical engineering converges on a finished artifact; software never settles, its substrate constantly renegotiated. SE is still young --- the NATO Conference dating from 1968 --- and applied unevenly. In sixty years that discipline has moved toward the classical engineering disciplines: formal methods, safety standards (DO-178C, IEC 62304, ISO 26262), Empirical Software Engineering, Model-Based Systems Engineering, product line engineering. Yet without it, mature or not, AI tools cannot resolve the problems the evidence lays bare. The two reinforce each other: SE \textit{supplies} what AI lacks, and AI \textit{accelerates} SE's systematic application and improvement.

\section{Why: The Structural Vacuum}\label{sec:vacuum}

The evidence and the practitioner voice reviewed in Section~1 reveal the vacuum; they do not explain it. To do so, we adopt a logical and epistemological approach: examining what AI coding agents structurally \textit{are} and, from that structure, deriving the roles human knowledge and judgment must necessarily play. The vacuum is not a performance gap: it is a structural condition.

Three properties of a large language model define how it operates and, together, constrain what it can and cannot do without external support.

\textbf{Probabilistic generation.} An LLM does not compute: it generates. Its output is a navigation over a probability distribution conditioned on training and context. And no digital artifact, generated or otherwise, can be validated against the world it represents without a reference external to the model.

\textbf{Agnosticism.} An LLM has absorbed patterns from vast corpora of code and text; three domains resist that absorption. Each has a reducible surface layer and a structurally irreducible core.

\textit{Methodological agnosticism.} The reducible surface is SE principles --- specification methods, architectural styles, testing frameworks --- which specialized models can absorb. The irreducible core is evaluative judgment: determining whether \textit{this} specification for \textit{this} system meets the bar requires a domain-grounded reference the model cannot generate. Asking the model to judge that adequacy itself does not help: it lacks that reference, which is difficult, if not impossible, to reify precisely each time.

\textit{Domain agnosticism.} The reducible surface is general domain knowledge (e.g., clinical constraints, regulatory requirements) accessible through RAG, fine-tuning, or domain training. The irreducible core is organization-specific knowledge: continuously renegotiated, semantically defined by the practice it describes, and permanently inconsistent in Balzer's sense. It can only be managed, never eliminated, and requires ongoing human mediation.

\textit{Evolutionary agnosticism.} The reducible surface is general evolution patterns --- what design choices constrain future change, what qualities matter in a given domain --- in principle trainable. The irreducible core is the project-specific trajectory: what \textit{this} system must become, given \textit{this} organization's constraints, does not exist until the software engineer reasons it into existence. This is a structural absence, not a deficiency of attention.

\textbf{Semantic statelessness.} Within a session, the model retains a working footprint: the context window gives it access to what was said, decided, and produced. Modern systems now add persistent memory across sessions, but as storage or log of prior exchanges --- not as a semantic model of accumulated know-how. The architectural decision made last Tuesday, the domain constraint introduced by the expert last week, the specification agreed upon last month: none becomes usable knowledge unless it has been reified as an artifact.

\section{Four Knowledge Levers, Three Actors}\label{sec:levers}

The vacuum has a precise structure. Filling it requires four distinct levers.

\textit{Methodological knowledge} --- how to build quality software --- is codified by SE as a discipline: fifty years of accumulated practice on specifications, architecture, testing, and process governance.

\textit{Domain knowledge} --- what the system must mean and do in its world --- is carried by domain experts: clinicians, operators, traders, and subject-matter specialists who understand the problem the software is built to solve.

\textit{Design choices} --- the project-specific decisions that translate into the product: architectural structures, component boundaries, trade-offs, interfaces. These are produced and owned by the software engineer, informed by both methodological and domain knowledge, and accumulate as the system grows.

\textit{Process choices} --- how the work is organized for this specific project: the lifecycle adopted, the criteria for agent autonomy, the review structure.

All four levers require \textbf{reification}: they must be materialized as persistent, accessible artifacts. Knowledge that exists only in someone's head does not exist for a stateless agent. Reification narrows the vacuum without closing it: the irreducible cores of Section~2 can be captured only in part, and what escapes the artifacts stays with the people accountable for the system.

Three actors do this work. \textit{Domain experts} supply semantic knowledge --- without them, SE structure is form without meaning. The \textit{AI agent} is the executor, powerful but without intrinsic reference for correctness or future orientation. The \textit{software engineer} is the active mediator between the two, and the custodian who keeps the system's evolutionary trajectory explicit over time.

One might object that these limits are temporary and that enough training data would let the agent learn without human intervention. But the three irreducible cores named in Section~2 are not gaps in training: evaluative judgment, organizational knowledge, and a system's future trajectory cannot be distilled from any corpus. The vacuum is structural, not historical.

\section{SE and AI}\label{sec:mechanisms}

The four levers and three actors just described operate through two distinct mechanisms by which SE fills the vacuum. The first is \textit{constitutive}: SE methodology embedded directly in the agent's design, so that it asks for specifications when they are missing, signals underspecified requirements, and structures its outputs according to SE principles before any human provides project-specific input. The second is \textit{operational}: the software engineer provides SE-structured artifacts --- specifications, architectural contexts, testing frameworks, domain constraints --- that give the agent a reference for correctness on each project. The two mechanisms map onto the two layers: constitutive addresses the reducible, operational addresses the irreducible core.

AI, in turn, helps SE close a knowing-doing gap it has long been unable to close on its own: formal requirements, architectural documentation, and systematic testing are demonstrably effective yet consistently underused. CASE tools in the 1980s and 1990s tried to close this gap, resented for their overhead. AI shifts the cost dynamic: reification, verification, and accountability (concepts SE has developed for decades) become cheaper to apply. Specifications, once expensive to produce and maintain outside critical domains, become byproducts of elicitation sessions and can be progressively formalized to reconnect with established SE practices. Tests, once written after the code to verify it, become the reference the agent generates against. Architecture Decision Records, once documentation debt, become the substrate the agent operates on. Practices once confined to critical domains can become ordinary, as the mechanical cost of applying the method drops below the threshold of abandonment.

The software engineer's role now operates on three axes: \textit{methodological} (providing the agent with sound SE structure); \textit{mediating} (eliciting organization-specific domain knowledge and structuring it as verifiable artifacts); and \textit{custodial} (owning the design and process choices that accumulate over the project's lifetime). Custody is, above all, accountability: agents can propose and execute decisions, but responsibility for them cannot be delegated. To govern an AI agent well, the engineer must know \textit{more} SE, not less. The pattern is visible at organizational scale: the DORA 2025 State of AI-assisted Software Development report, based on nearly five thousand professionals worldwide, finds that AI adoption amplifies practices already present. Teams with mature engineering foundations convert AI into delivery throughput; teams without them find AI accelerating instability rather than productivity.

\section{In Practice: Artifacts and Workflows}\label{sec:practice}

Sections~\ref{sec:levers} and~\ref{sec:mechanisms} established that the four knowledge levers must be reified as artifacts, and that AI has made their reification sustainable beyond the critical domains that could once justify its cost. The artifacts named below are largely independent of the software lifecycle adopted; the list is not exhaustive.

\textit{Specifications} (the primary operational artifact of spec-driven development) supply the model with the criterion of adequacy it cannot generate for itself. \textit{Architecture Decision Records} encode the design choices that accumulate over the system's life, along with the \textit{why} and the intended trajectory that no session context preserves. \textit{Domain constraint records} reify the organization-specific knowledge elicited from experts: the artifact form of what is otherwise permanently inconsistent and never articulated. \textit{Test frameworks structured for agent input} externalize the reference of correctness the agent has no internal access to, becoming the agent's proving ground and the engineer's acceptance criterion. \textit{Process definitions instantiated per project} make process choices executable rather than aspirational: the criteria for agent autonomy, the boundaries of automated action, the review structure. \textit{Design rationale records}, finally, hold the custodial layer of coherence across accumulated decisions, distinguishing incidental patches from principled changes.

A natural question is whether AI changes development workflows. AI does not prescribe one. Any lifecycle a team adopts (waterfall, agile, spec-driven, or incremental refinement of existing practices) can be supported by an AI engine, provided the artifacts are present and continuously maintained. The ``loop and graph engineering'' now celebrated in agentic development revisits iterative, incremental, or more structured development --- long practiced in SE --- now driven by an AI executor: AI empowering an established method. The ambition to drive the software process through such an engine is not new. Through the 1990s and 2000s, research on Process Modeling Languages and Process-centered Software Engineering Environments attempted to govern the software process through deterministic formal models. A 2014 retrospective concluded that formal process automation could succeed only in narrow technical phases such as code generation, testing, and deployment~\cite{fuggetta2014}. Deterministic execution cannot govern an activity whose normal state is inconsistency: it either blocks or forces premature resolution that erases information. AI resolves the problem differently: a probabilistic engine does not require constraint satisfaction, navigates incomplete specs, and surfaces inconsistencies as outputs for human judgment. The intent behind PSEE research is not abandoned. Where deterministic execution could not, probabilistic delivery opens a new direction.

\section{Conclusions}

SE is not a discipline in decline. It is a discipline whose moment has arrived --- the one that defines the structured space in which AI agents can be trusted. And the software engineer is not a professional in decline either --- but the one who carries that knowledge into practice: who defines what the agent navigates by, who elicits the domain meaning that no corpus can supply, and who owns the trajectory of the system over time.

Three implications follow.

\textbf{Reframe the question.} The dominant debate asks whether AI will replace software engineers. The right question is different: what structured knowledge must govern AI agents to produce software that is correct, trustworthy, and built to last? Answering it requires SE and software engineers, not their abandonment.

\textbf{Redirect the investment.} Specifications, domain elicitation, architectural governance, and process discipline are not overhead to minimize --- they are the structured space within which AI agents can be trusted to operate with persistent and coherent memory. Organizations that treat them as optional will pay the cost in correctness, maintainability, and accountability.

\textbf{Revalue the discipline.} SE education, research, and practice are not legacies to be displaced by AI fluency. They are prerequisites for AI exploitation and governance. A generation of developers trained only in prompting --- without methodological knowledge, domain elicitation skill, or architectural judgment --- will reproduce at scale the failure modes this paper describes. And without early-career hiring, as Russinovich and Hanselman warn in these pages, the profession's talent pipeline will collapse.

\begin{acks}
The author thanks Damian Tamburri for his feedback on different revisions of this paper. Claude (Anthropic) was used as an AI writing assistant during drafting and revision. All arguments, claims, and editorial decisions are the author's own.
\end{acks}


\begin{thebibliography}{10}

\bibitem{agarwal2026}
S. Agarwal, H. He, and B. Vasilescu. AI IDEs or Autonomous Agents? Measuring the Impact of Coding Agents on Software Development. In \textit{Proc. 23rd Int. Conf. on Mining Software Repositories (MSR '26)}, Rio de Janeiro, Brazil, 2026, pp. 857--862. doi:10.1145/3793302.3793589.

\bibitem{balzer1991}
R. Balzer. Tolerating Inconsistency. In \textit{Proc. 13th Int. Conf. on Software Engineering (ICSE 1991)}, Austin, TX, 1991, pp. 158--165.

\bibitem{becker2025}
J. Becker, N. Rush, E. Barnes, and D. Rein. Measuring the Impact of Early--2025 AI on Experienced Open-Source Developer Productivity. arXiv:2507.09089, July 2025.

\bibitem{cui2025}
K. Z. Cui, M. Demirer, S. Jaffe, L. Musolff, S. Peng, and T. Salz. The Effects of Generative AI on High-Skilled Work: Evidence from Three Field Experiments with Software Developers. \textit{Management Science}, 2026. doi:10.1287/mnsc.2025.00535.

\bibitem{daniotti2026}
S. Daniotti, J. Wachs, X. Feng, and F. Neffke. Who Is Using AI to Code? Global Diffusion and Impact of Generative AI. \textit{Science}, 391(6787):831--835, 2026. doi:10.1126/science.adz9311.

\bibitem{fuggetta2014}
A. Fuggetta and E. Di Nitto. Software Process. In \textit{Proc. Future of Software Engineering (FOSE'14)}, Hyderabad, India, 2014, pp. 1--14.

\bibitem{he2026}
H. He, C. Miller, S. Agarwal, C. K\"{a}stner, and B. Vasilescu. Speed at the Cost of Quality: How Cursor AI Increases Short-Term Velocity and Long-Term Complexity in Open-Source Projects. In \textit{Proc. 23rd Int. Conf. on Mining Software Repositories (MSR '26)}, Rio de Janeiro, Brazil, 2026, pp. 181--193. doi:10.1145/3793302.3793349.

\bibitem{huang2025}
R. Huang, A. Reyna, S. Lerner, H. Xia, and B. Hempel. Professional Software Developers Don't Vibe, They Control: AI Agent Use for Coding in 2025. arXiv:2512.14012v2, Aug. 2026.

\bibitem{meyer2026}
B. Meyer. Artificial Intelligence for Software Engineering: From Probable to Provable. \textit{Communications of the ACM}, 69(6):46--49, June 2026. doi:10.1145/3773295.

\bibitem{xu2025}
F. Xu, P. K. Medappa, M. M. Tunc, M. Vroegindeweij, and J. C. Fransoo. AI-Assisted Programming Decreases the Productivity of Experienced Developers by Increasing the Technical Debt and Maintenance Burden. arXiv:2510.10165v3, Jan. 2026.

\end{thebibliography}
\end{document}